\documentclass{PoS}

\def\Journal#1#2#3#4{{#1} {\bf #2}, #3 (#4)}

\title{
  Transverse momentum spectra of identified charged hadrons with the
  ALICE detector in Pb--Pb collisions at $\sqrt{s_{NN}}$~=~2.76~TeV
}

\ShortTitle{
  Transverse momentum spectra of hadrons with ALICE
}

\author{
  \speaker{
    Roberto Preghenella
  } 
  (for the ALICE Collaboration)\\
  Centro Studi e Ricerche e Museo Storico della Fisica ``Enrico Fermi'',
  Rome, Italy\\
  Dipartimento di Fisica dell'Universit\`a and Sezione INFN, Bologna,
  Italy\\
  E-mail: \email{preghenella@bo.infn.it}
}

\abstract{
  The measurement of identified charged-hadron production at mid-rapidity
  ($\left|y\right|~<~0.5$) performed with the ALICE experiment is
  presented for Pb--Pb collisions at $\sqrt{s_{\rm NN}}~=~\rm~2.76~TeV$. The
  transverse momentum spectra of $\pi^{\pm}$, $K^{\pm}$, $p$ and $\bar{p}$ are
  measured from 100~MeV/c up to 3~GeV/c for pions, from 200~MeV/c up to
  2~GeV/c for kaons and from 300~MeV/c up to 3~GeV/c for protons and
  antiprotons using the \emph{dE/dx} and the \emph{time-of-flight}
  particle-identification techniques. Preliminary results on charged hadron
  production yields and particle ratios are reported as a function of $p_{T}$
  and collision centrality. Finally, the results are discussed in terms of
  hydrodynamics-inspired models and compared with published RHIC data in Au--Au 
  collisions at $\sqrt{s_{\rm NN}}~=~\rm~200~GeV$ and predictions for the LHC.
}

\begin{document}

\section{Introduction}\label{sec:introduction}
ALICE (A Large Ion Collider Experiment) is a general-purpose heavy-ion
experiment at the CERN LHC (Large Hadron Collider) aimed at studying the
physics of strongly-interacting matter and the quark--gluon plasma. A unique
design has been adopted for the ALICE detector to fulfill tracking and
particle-identification requirements~\cite{ref:ALICEperf}. Thanks to these features the experiment
is able to identify charged hadrons with momenta from about 0.1~GeV/c and up
to a few GeV/c by combining different detecting systems, as discussed in
Section~\ref{sec:pid}.

The hot and dense matter produced in ultrarelativistic heavy-ion collisions
evolves through different phases to a freeze-out state where strong interactions
among the hadrons stop. Since produced hadrons carry information about the
evolution of the system, the measurement of the tranverse momentum
distributions and yields of identified hadrons is essential to understand the
global properties and dynamics of the later stages. Results on
charged-hadron spectra and yields at mid-rapidity are presented in
Section~\ref{sec:results} for Pb--Pb collisions at
$\sqrt{s_{\rm NN}}~=~\rm~2.76~TeV$. 

\section{Particle identification}\label{sec:pid}
In this section the particle-identification (PID) detectors relevant for this
analysis are briefly discussed, namely the \emph{Inner Tracking System} (ITS),
the \emph{Time-Projection Chamber} (TPC) and the \emph{Time-Of-Flight} detector
(TOF). A detailed review of the ALICE detector and of its PID capabilities can
be found in~\cite{ref:ALICEperf}. The ITS is a six-layer silicon detector located at radii between 4 and 43
cm. Four of the six layers provide $dE/dx$ measurements and are used for
particle identification in the non-relativistic ($1/\beta^2$)
region. Moreover, using the ITS as a standalone tracker enables one to
reconstruct and identify low-momentum particles not reaching the main tracking
systems. The TPC is the main central-barrel tracking detector of ALICE and
provides three-dimensional hit information and specific energy-loss
measurements with up to 159 samples. With the measured particle momentum and
$\langle dE/dx \rangle$ the particle type can be determined by comparing the
measurements against the Bethe-Bloch expectation. The TOF detector is a
large-area array of Multigap Resistive Plate Chambers (MRPC) and covers the central
pseudorapidity region ($\left| \eta \right| <$~0.9, full azimuth). Particle
identification is performed by matching momentum and trajectory-length
measurements performed by the tracking system with the time-of-flight
information provided by the TOF system. The total time-of-flight resolution is
about 85 ps in Pb--Pb collisions and it is determined by the time resolution
of the detector itself and by the start-time resolution.

\section{Results}\label{sec:results}
\begin{figure}[t]
  \centering
  \begin{minipage}[t]{0.49\linewidth}
    \centering
    \includegraphics[width=0.9\textwidth]{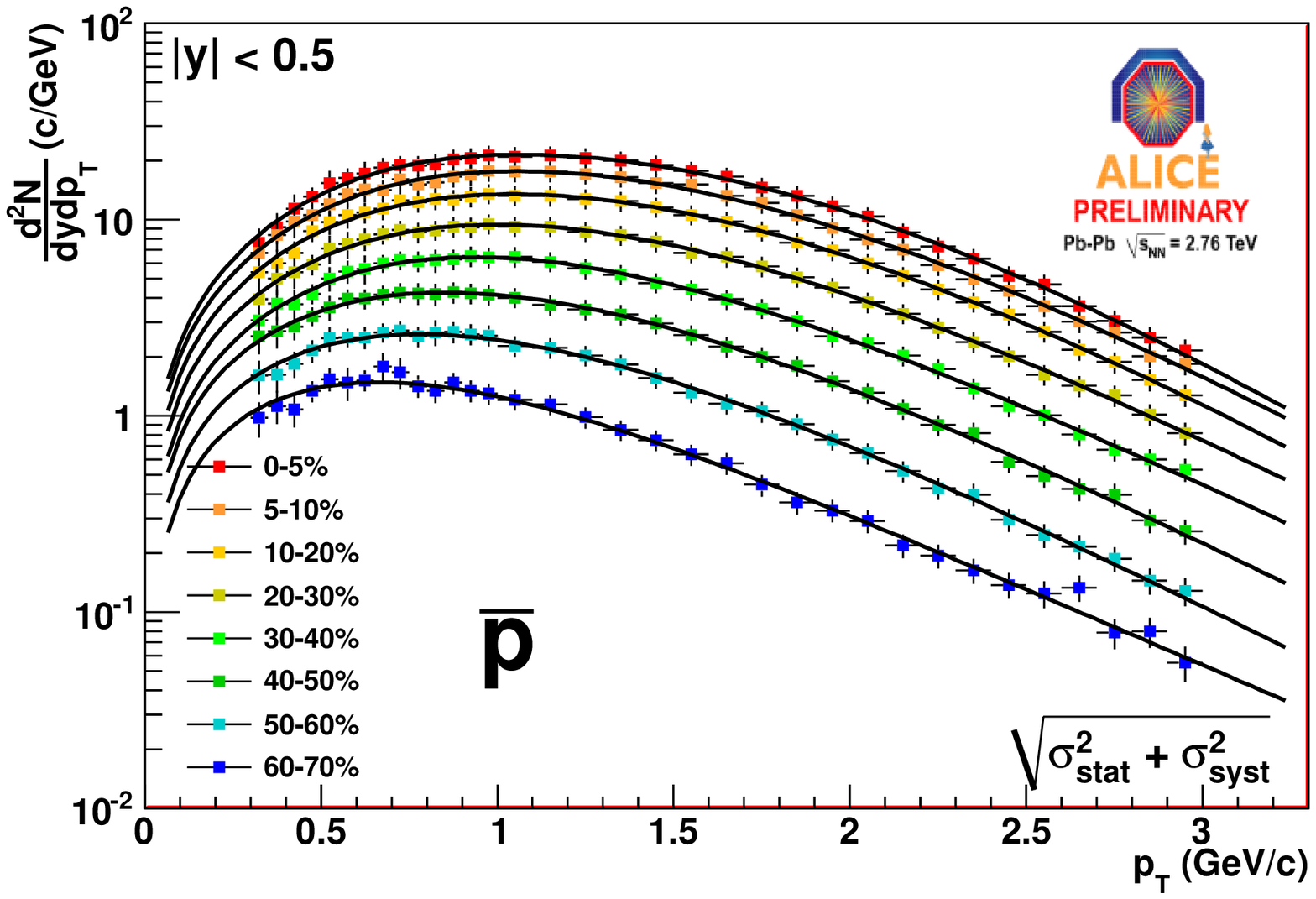}
    \vspace{-5mm}
    \caption{Transverse momentum spectra of primary $\bar{p}$ and
      corresponding fits in several centrality classes.}
    \label{fig:protonspectra}
  \end{minipage}
  \hfill
  \begin{minipage}[t]{0.49\linewidth}
    \centering
    \includegraphics[width=0.9\linewidth]{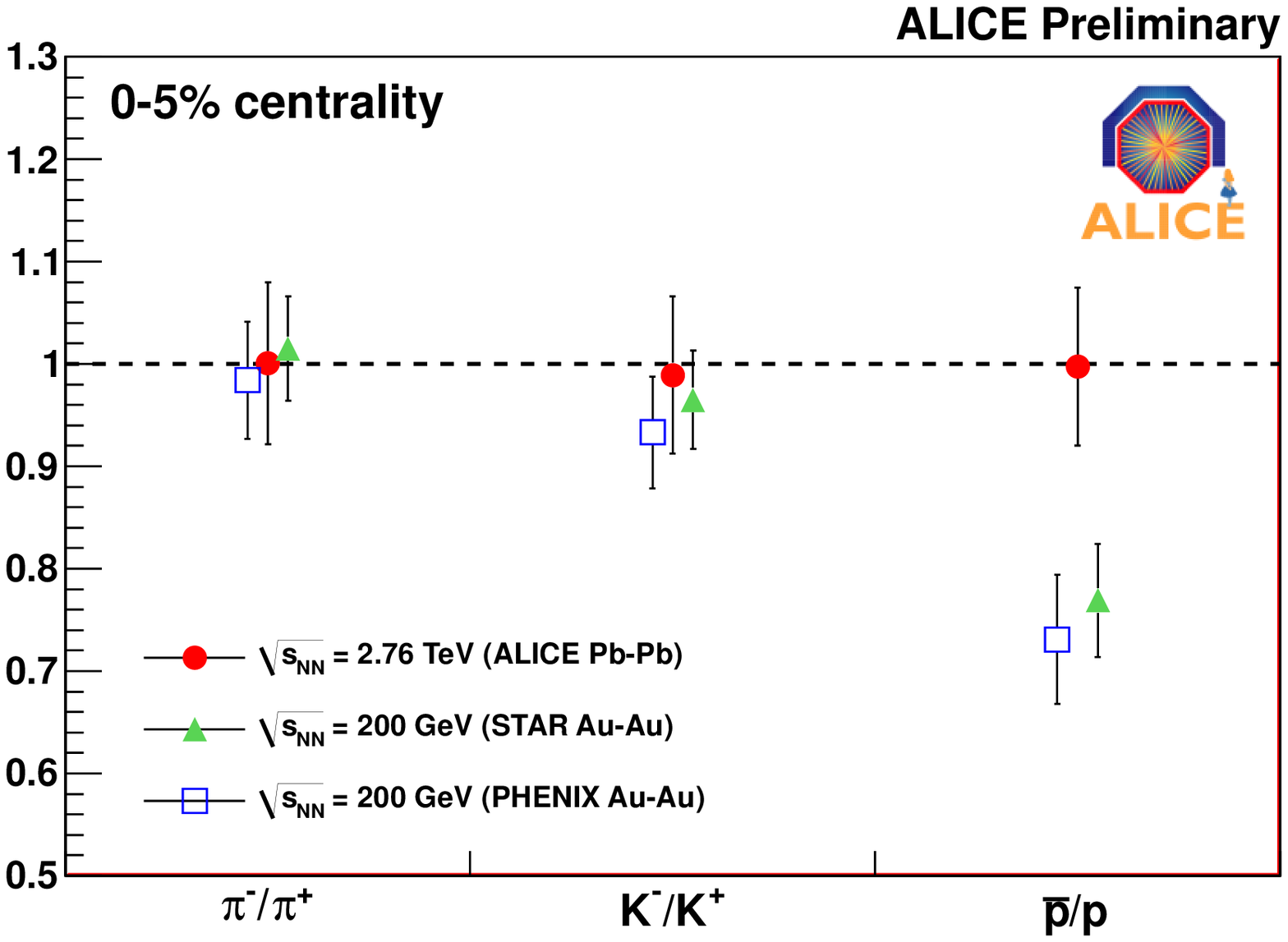}
    \vspace{-5mm}
    \caption{Antiparticle/particle production ratios in the 0-5\% most central
      collisions.} 
    \label{fig:mub}
  \end{minipage}
  \begin{minipage}[t]{0.49\linewidth}
    \centering
    \includegraphics[width=0.9\textwidth]{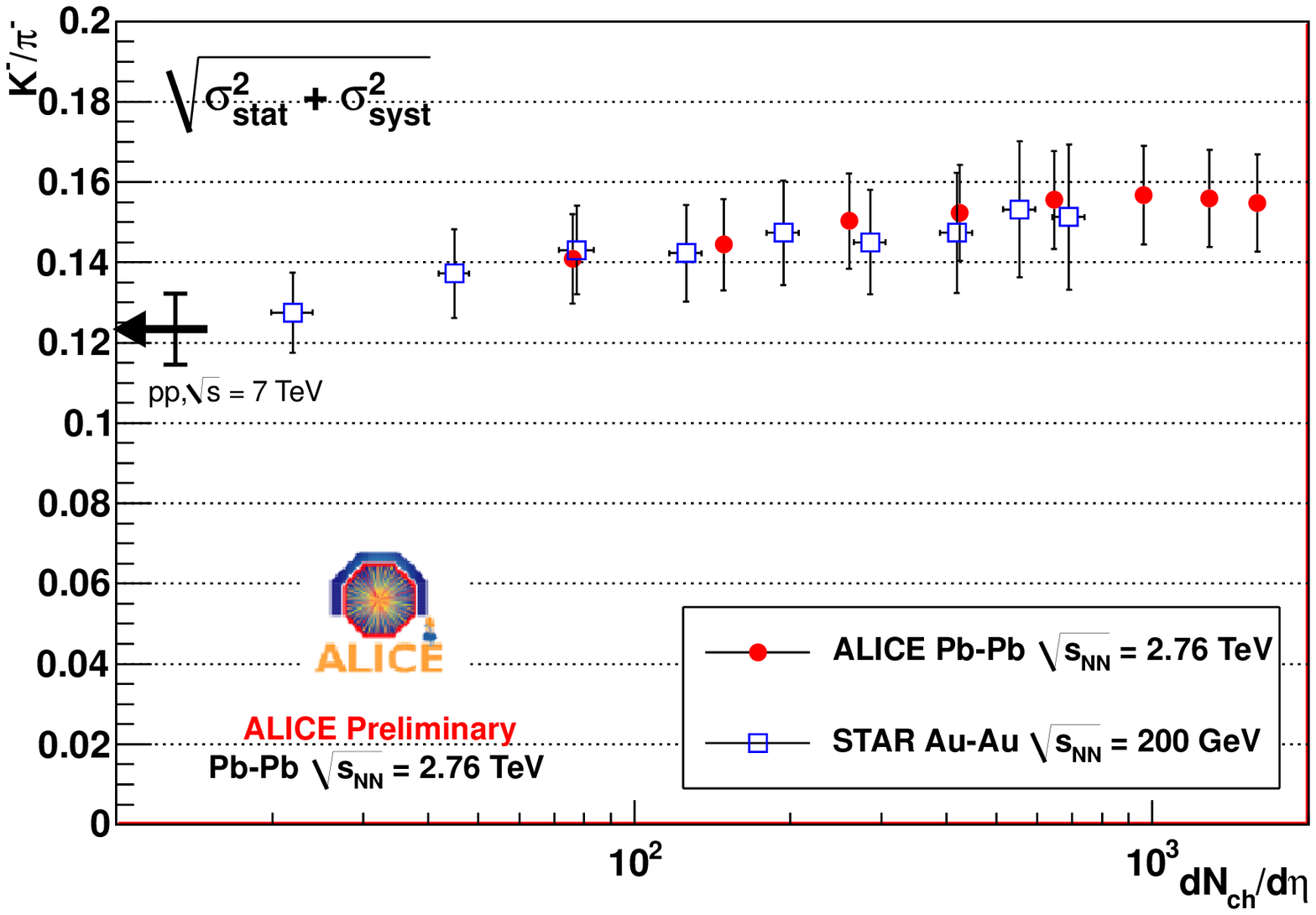}
    \vspace{-5mm}
    \caption{$K^{-}/\pi^{-}$ production ratios as a function of
      $dN_{ch}/d\eta$ compared to RHIC data.}
    \label{fig:kapi}
  \end{minipage}
  \hfill
  \begin{minipage}[t]{0.49\linewidth}
    \centering
    \includegraphics[width=0.9\textwidth]{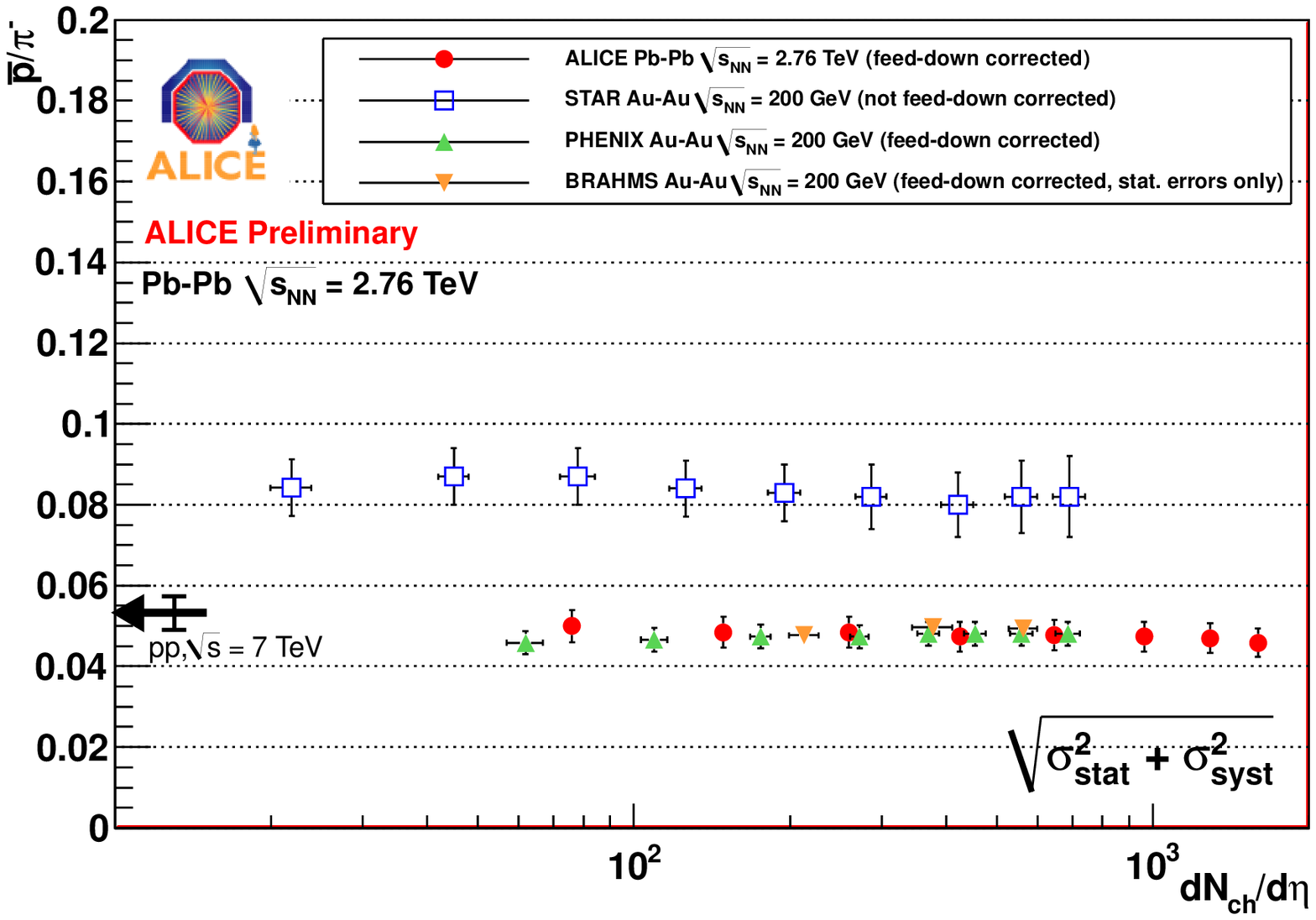}
    \vspace{-5mm}
    \caption{$\bar{p}/\pi^{-}$ production ratios as a function of
      $dN_{ch}/d\eta$ compared to RHIC data.} 
    \label{fig:prpi}
  \end{minipage}
\end{figure}
The transverse momentum spectra of primary $\pi^{\pm}$, $K^{\pm}$, $p$ and
$\bar{p}$ are measured at mid-rapidity ($\left|y\right|~<~0.5$) combining the
techniques and detectors described in Section~\ref{sec:pid}. Primary particles
are defined as prompt particles produced in the collision and all decay
products, except products from weak decay of strange particles. The
contribution from the feed-down of weakly-decaying particles to $\pi^{\pm}$,
$p$ and $\bar{p}$ and from protons from material are subtracted by fitting the
data using Monte Carlo templates of the DCA\footnote{Distance of Closest
  Approach to the reconstructed primary vertex.} distributions. Hadron spectra are
measured in several centrality classes (see~\cite{ref:ALICEpbpb} for details
on centrality selection) from 100~MeV/c up to 3~GeV/c for pions, from
200~MeV/c up to 2~GeV/c for kaons and from 300~MeV/c up to 3~GeV/c for
protons and antiprotons. Individual fits to the data are performed following a
blast-wave parameterization~\cite{ref:blastwave} to extrapolate the
spectra outside the measured $p_{T}$ range. The measured spectra and corresponding fits are shown in
Figure~\ref{fig:protonspectra} for primary $\bar{p}$. Average transverse
momenta $\langle p_{T} \rangle$ and integrated 
production yields $dN/dy$ are obtained using the measured data points and the
extrapolation. 

Antiparticle/particle integrated production ratios are observed to be
consistent with unity for all particle species in all centralities suggesting
that the baryo-chemical potential $\mu_{B}$ is close to zero as expected at LHC
energies. Figure~\ref{fig:mub} compares ALICE results with
RHIC at  data in Au--Au 
collisions at $\sqrt{s_{\rm NN}}$ = 200 GeV~\cite{ref:RHIC} for the 0-5\%
most central collisions. The $p_{T}$-integrated $K^{-}/\pi^{-}$ and $\bar{p}/\pi^{-}$ ratios are shown in
Figure~\ref{fig:kapi} and~\ref{fig:prpi} as a function of the charged-particle
density $dN_{ch}/d\eta$~\cite{ref:ALICEpbpb} and are compared with RHIC data at 
$\sqrt{s_{\rm NN}}$~=~200~GeV and
ALICE proton-proton results at
$\sqrt{s}$~=~7~TeV~\cite{ref:MarekQM}. $K^{-}/\pi^{-}$ production nicely 
follows the trend measured by STAR. $\bar{p}/\pi^{-}$ results are similar to
previous measurements performed by PHENIX and BRAHMS where the definition of
the proton sample is close to ours (proton measurements reported by STAR are inclusive). Finally, the $\bar{p}/\pi^{-}$ ratio measured at the LHC ($\sim$~0.05) is
significantly lower that the value expected from statistical model predictions
($\sim$~0.07-0.09) with a chemical freeze-out temperature of $T_{ch} =
160-170$~MeV at the LHC~\cite{ref:statmodels}.

\begin{figure}[t]
  \centering
  \begin{minipage}[t]{0.49\linewidth}
    \centering
    \includegraphics[width=0.9\textwidth]{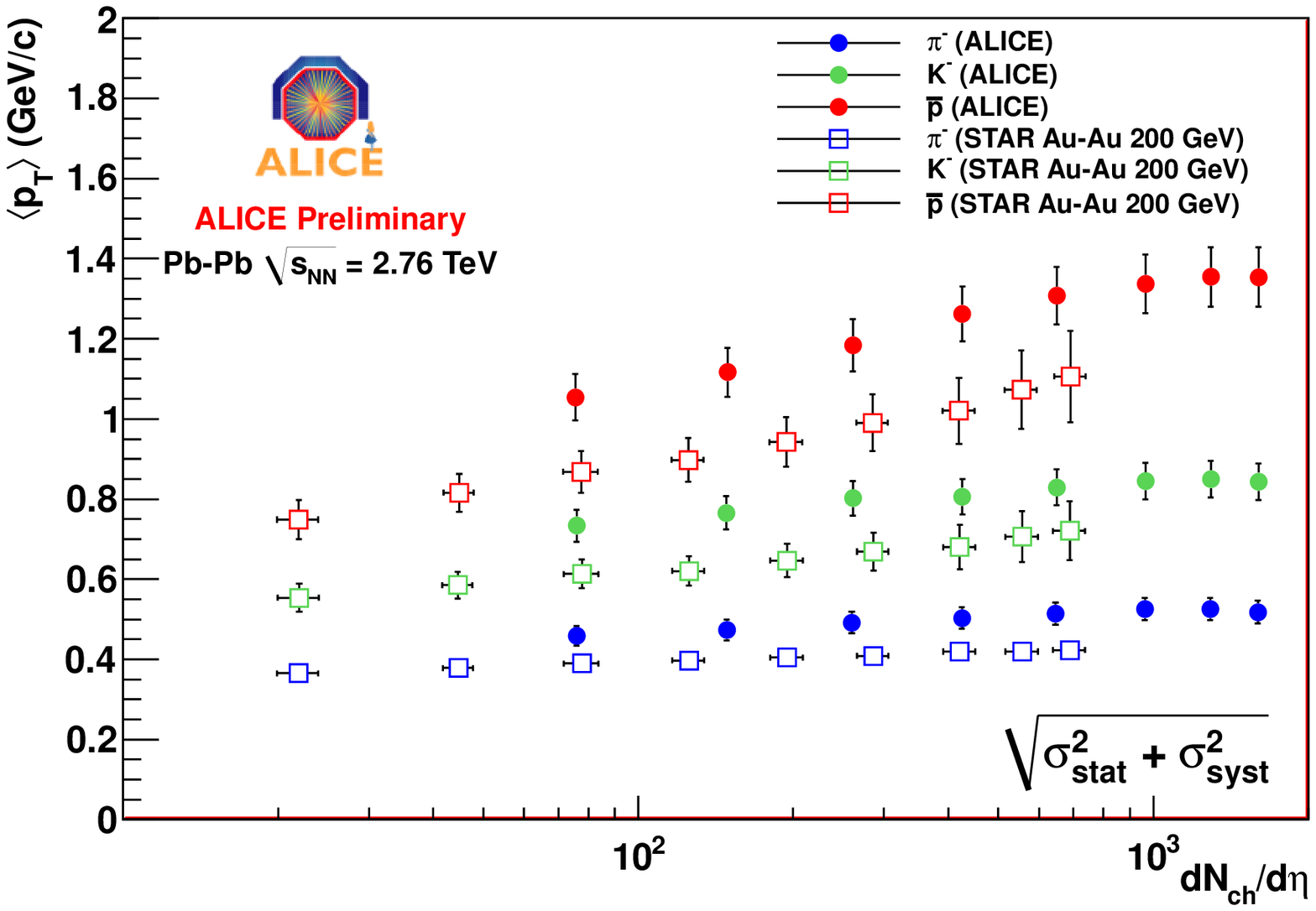}
    \vspace{-5mm}
    \caption{Hadron $\langle p_{T} \rangle$ as a function of the
      charged-particle density $dN_{ch}/d\eta$.}
    \label{fig:meanpt}
  \end{minipage}
  \hfill
  \begin{minipage}[t]{0.49\linewidth}
    \centering
    \includegraphics[width=0.9\textwidth]{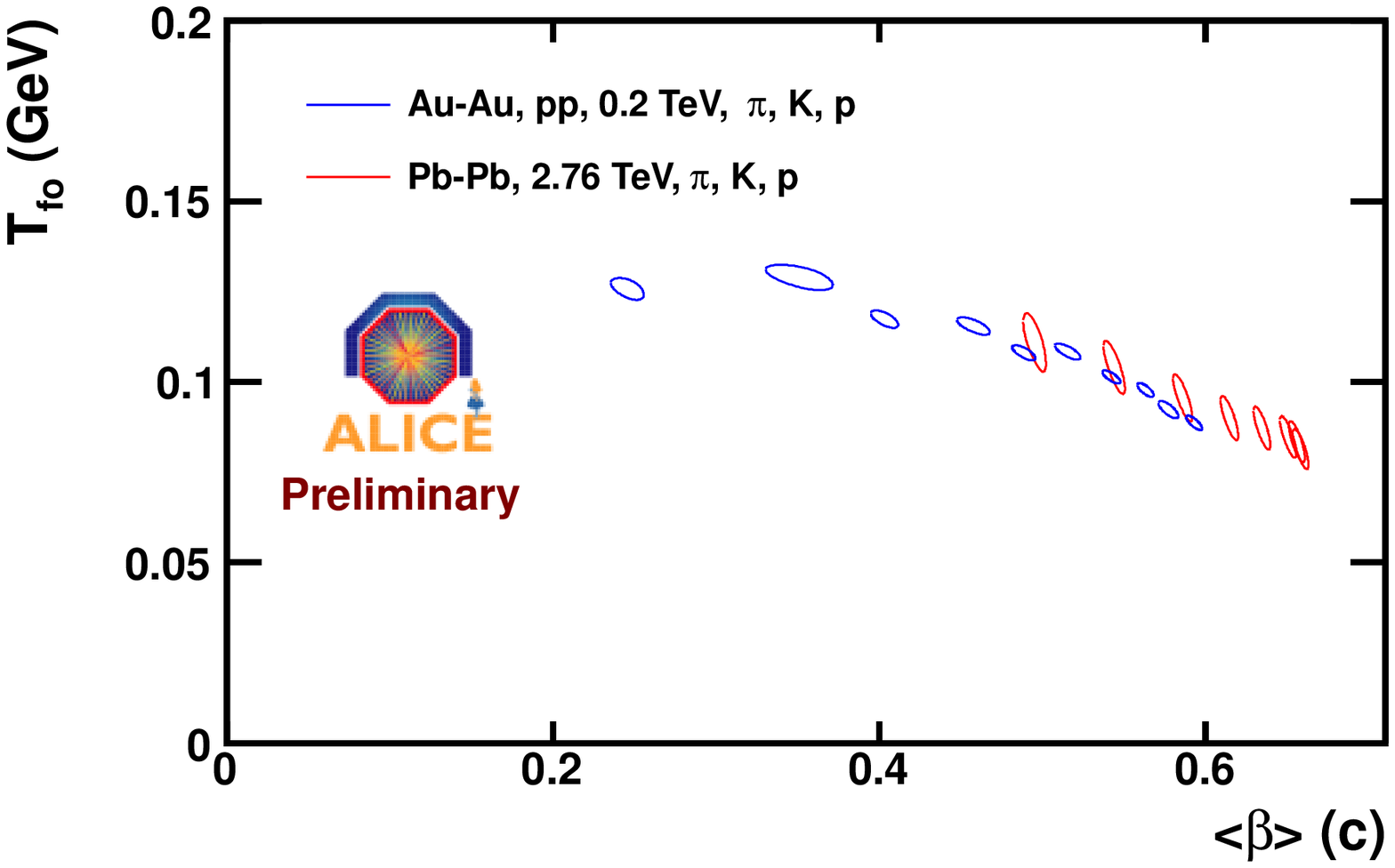}
    \vspace{-5mm}
    \caption{Thermal freeze-out parameters $T_{fo}$ and $\langle \beta \rangle$
      from combined blast-wave fits.}
    \label{fig:blastwave}
  \end{minipage}
\end{figure}

The measured hadron $\langle p_{T} \rangle$'s are shown in
Figure~\ref{fig:meanpt} as a function of $dN_{ch}/d\eta$ for $\pi^{-}$, $K^{-}$ and $\bar{p}$ and
are compared to STAR results in Au--Au collisions at
$\sqrt{s_{\rm NN}}$~=~200~GeV. The spectra are observed to be
harder that at RHIC for similar $dN_{ch}/d\eta$. A detailed study of the
spectral shapes has been done in order to give a quantitative estimate of
the thermal freeze-out temperature $T_{fo}$ and the average transverse flow
$\langle \beta \rangle$. A combined blast-wave fit of the spectra has
been performed in the ranges 0.3-1.0~GeV/c, 0.2-1.5~GeV/c and 0.3-3.0~GeV/c
for pions, kaons and protons respectively. While the $T_{fo}$ parameter is
slightly sensitive to the pion fit range because of feed-down of
resonances\footnote{This effect will be investigated in details in the
  future.} the transverse flow $\langle \beta \rangle$ measurement is not,
being dominated by the proton spectral shape. The results obtained on the
thermal freeze-out properties in different centrality bins are compared with
similar measurements performed by the STAR Collaboration at lower energies in
Figure~\ref{fig:blastwave}. A stronger radial flow is observed with respect to
RHIC, being about 10\% larger in the most central collisions at the LHC.

\section{Conclusions}\label{sec:conclusions}
The tranverse momentum spectra of $\pi^{\pm}$, $K^{\pm}$, $p$ and $\bar{p}$ have
been measured with ALICE in Pb--Pb collisions at $\sqrt{s_{\rm NN}}$~=~2.76~TeV,
demonstrating the excellent PID capabilities of the
experiment. Antiparticle/particle production ratios are consistent with unity as  
expected at LHC energies. $\bar{p}/\pi^{-}$ integrated ratio is significantly
lower than statistical model predictions with a chemical freeze-out temperature $T_{ch} =
160-170$~MeV. The average transverse momenta and the spectral shapes indicate
a $\sim$10\% stronger radial flow than at RHIC energies.

\end{document}